\documentclass[prd,twocolumn,aps,showpacs,nofootinbib,nobibnotes,superscriptaddress,preprintnumbers]{revtex4}
\usepackage{epsfig}
\usepackage{graphics}
\usepackage{bm}
\usepackage{color}
\usepackage{dcolumn}   
\usepackage[spanish,english]{babel}
\usepackage{bm}     
\usepackage{bbm}       
\usepackage{amssymb}  
\usepackage{amsmath}
\usepackage{latexsym}
\usepackage{float}
\usepackage{ifthen}
\usepackage{caption,subfig}
\usepackage{enumerate}
\usepackage{url}
\usepackage{caption,subfig}
\usepackage{amsopn}
\usepackage{hyperref}

\usepackage{amsfonts}
\usepackage{multirow}
\usepackage{array}
\usepackage{booktabs}
\usepackage{rotating}

\usepackage{ulem}
\def\clap#1{\hbox to 0pt{\hss#1\hss}}

\def\({\left(}
\def\){\right)}
\def\[{\left[}
\def\]{\right]}
\def\bea{\begin{eqnarray}}
\def\eea{\end{eqnarray}}
\def\be{\begin{equation}}
\def\ee{\end{equation}}
\def\ba{\begin{eqnarray}}
\def\ea{\end{eqnarray}}
\def\beq{\begin{eqnarray}}
\def\eeq{\end{eqnarray}}

\def\d{\mathrm{d}}
\newcommand{\cs}{c_s}

\newcommand{\M}{\mathcal{M}}
\newcommand{\N}{\mathcal{N}}
\newcommand{\Pm}{{\mathcal P}}
\newcommand{\Qm}{{\mathcal Q}}

\newcommand{\Ft}{\tilde{F}}

\newcommand{\LL}{\mathcal{L}}
\newcommand{\Tr}{{\rm Tr}}

\newcommand{\Id}{\mathbbm 1}

\def\cs{c_{\rm s}}

\def\be{\begin{equation}}
\def\ee{\end{equation}}
\def\ba{\begin{eqnarray}}
\def\ea{\end{eqnarray}}
\def\beq{\begin{eqnarray}}
\def\eeq{\end{eqnarray}}

\def\d{\mathrm{d}}

\def\K{{\cal K}}

\def\L*{{\cal L}_*}
\def\L{\mathcal{L}}
\def\({\left(}
\def\){\right)}

\def\<{\langle}
\def\>{\rangle}

\def\cs2{c_{s}^{2}}

\def\be{\begin{equation}}
\def\ee{\end{equation}}
\def\ba{\begin{eqnarray}}
\def\ea{\end{eqnarray}}
\def\beq{\begin{eqnarray}}
\def\eeq{\end{eqnarray}}
\def\M{\mathcal{M}}

\def\d{\mathrm{d}}

\def\K{{\cal K}}

\def\L*{{\cal L}_*}
\def\L{\mathcal{L}}
\def\({\left(}
\def\){\right)}

\def\<{\langle}
\def\>{\rangle}

\def\Tr{{\rm Tr}}

\begin{document}

\title{Generalized multi-Proca fields}

\date{\today,~ $ $}

\author{Jose Beltr\'an Jim\'enez} \email{jose.beltran@cpt.univ-mrs.fr}
\affiliation{Aix Marseille Univ, Universit\'e de Toulon, CNRS, CPT, Marseille, France}

\author{Lavinia Heisenberg} \email{lavinia.heisenberg@eth-its.ethz.ch}
\affiliation{Institute for Theoretical Studies, ETH Zurich, 
\\ Clausiusstrasse 47, 8092 Zurich, Switzerland}

\date{\today}

\begin{abstract}
We extend previous results on healthy derivative self-interactions for a Proca field to the case of a set of massive vector fields. We obtain non-gauge invariant derivative self-interactions for the vector fields that maintain the appropriate number of propagating degrees of freedom. In view of the potential cosmological applications, we restrict to interactions with an internal rotational symmetry.
We provide a systematical construction order by order in derivatives of the fields and making use of the antisymmetric Levi-Civita tensor. We then compare with the one single vector field case and show that the interactions can be broadly divided into two groups, namely the ones obtained from a direct extension of the generalized Proca terms and genuine multi-Proca interactions with no correspondence in the single Proca case. We also discuss the curved spacetime version of the interactions to include the necessary non-minimal couplings to gravity. Finally, we explore the cosmological applications and show that there are three different vector fields configurations giving rise to isotropic solutions. Two of them have already been considered in the literature and the third one, representing a combination of the first two, is new and offers unexplored cosmological scenarios.

\end{abstract}


\maketitle

\section{Introduction}
The accelerated expansion of the universe discovered almost two decades ago still remains a challenging puzzle for modern cosmology. Assuming General Relativity as the appropriate theory describing the gravitational interaction on cosmological scales, the cosmic acceleration can be  accounted for by simply including a cosmological constant. However, its required value in agreement  with observations turns out to be tiny as compared to the expected {\it natural} value and this discord puts on trial our theoretical understanding of gravity and the standard techniques of quantum field theory \cite{Weinberg:1988cp}. This problem has triggered a plethora of attempts to modify gravity on  large scales and most of them eventually unfold in the form of additional scalar fields, which can then be used as a condensate whose energy density can drive the accelerated expansion of the universe. Some modified gravity scenarios resorted to braneworld models with extra-dimensions as possible mechanisms to generate acceleration and/or alleviate the hierarchy problem, being the DGP  model \cite{Dvali:2000hr} a paradigmatic example. In this model, the effective scalar field describing the vibrations of the brane presents interesting features among which we can mention interactions involving second derivatives of the scalar field, which nevertheless lead to second order field equations so that the Ostrogradski instability is avoided, and a Galilean symmetry allowing for constant shift not only in the field itself, but also in its gradient.  These properties were then generalized to find the most general Lagrangian sharing such features \cite{Nicolis:2008in} and are known as Galileon interactions. A nice property of these interactions is their radiative stability under quantum corrections \cite{quantum_corrections}, even if they fail to tackle the cosmological constant problem. The covariantization of these Galileon interactions to include gravity requires the introduction of non-minimal couplings in order to maintain the second order nature of the equations of motion and this led to the rediscovery of the now so-called Horndeski Lagrangians, which are the most general scalar-tensor theories  leading to second order equations of motion \cite{Horndeski:1974wa}. Further developments showed that it is in fact possible to build more general scalar-tensor theories with higher order equations of motion, but still without additional propagating degrees of freedom and, therefore, avoiding the Ostrogradski instability \cite{beyondH}.
 
Along the lines of constructing consistent theories for scalar-tensor interactions, one can try to build analogous consistent theories for a vector field. Interestingly, very much like the Galilelon interactions can be elegantly obtained from geometrical constructions in higher dimensions \cite{deRham:2010eu,VanAcoleyen:2011mj}, it is possible to obtain vector Galileon interactions within the framework of (generalized) Weyl geometries \cite{Jimenez:2014rna,Jimenez:2015fva}.
For a massive vector field one can indeed construct non-gauge invariant derivative self-interactions of the vector field with the requirement that only three degrees of freedom propagate, as it corresponds to a massive vector field. The resulting theory is composed by the generalized Proca interactions which guarantee having second order equations of motion and the desired 3 polarizations for the vector \cite{genProca,JbJLH}. Not surprisingly and in a similar way to the Horndeski lagrangians for a scalar field, the generalized Proca interactions can also be further extended to the case of more general vector-tensor interactions with higher order equations of motion, but still propagating three polarizations \cite{beyondgenProca}.

The goal of this work is to extend the generalized Proca interactions to the case of several interacting vector fields to obtain a multi-Proca version of the healthy non-gauge invariant derivative interactions. A similar extension has also been pursued for the case of scalar Galileon interactions, resulting in the so-called multi-Galileon theories \cite{multigalileon}. We will apply a construction scheme taking advantage of the symmetries of the Levi-Civita tensor in the same spirit as the one applied to the single Proca field in \cite{JbJLH}. 
For this purpose, we will go order by order in derivatives of the vector fields and construct the interactions to guarantee that the temporal components of the vector fields do not propagate and, thus,  giving rise to healthy interactions. Interactions for a generalized $SU(2)$ Proca field has been considered in \cite{Allys:2016kbq}. That work has a similar goal to ours, but using a different approach and limiting to interactions with up to six Lorentz indices. At the coincident orders, our interactions agree with theirs. Furthermore, our different systematical procedure allows us to construct interactions which are beyond the orders considered in \cite{Allys:2016kbq}.

The paper will be organized as follows. We will start by very briefly reviewing non-abelian gauge theories. In section \ref{Sec:SystConst} we will proceed to the systematical construction of the interactions, ending with a summary of all the interactions. Section \ref{Sec:RelProca} will be devoted to making a comparison between the obtained interactions and those present in the single vector field case. Furthermore, this will allow us to identify some of the required non-minimal couplings to extend the interactions to curved spacetime. Finally, in section \ref{Sec:CosmApp} we will discuss the possible configurations for the vector fields that will allow for isotropic cosmological solutions and illustrate it with a simple example. In section \ref{Sec:Conc} we discuss our main findings.

Internal group and Lorentz (spacetime) indices will be denoted by latin $a,b,c,\dots$ and greek $\alpha,\beta,\gamma,\dots$ letters respectively. We will use the mostly plus signature for the spacetime metric. The dual of an antisymmetric tensor $F_{\mu\nu}$ is defined as $\Ft^{\mu\nu}\equiv \frac 12 \epsilon^{\mu\nu\rho\sigma}F_{\rho\sigma}$. We define symmetrization and antisymmetrization as $T_{(\mu\nu)}=T_{\mu\nu}+T_{\nu\mu}$ and $T_{[\mu\nu]}=T_{\mu\nu}-T_{\nu\mu}$ respectively.
\section{Non-abelian gauge field}
Before proceeding to the construction of the derivative self-interactions for a set of massive vector fields, it will be convenient to briefly review the properties of interacting massless vector fields. It is known that consistency of the interactions for the massless vector fields leads to the full non-abelian gauge structure of Yang-Mills theories. Alternatively, one can start with the Lagrangian for a set of massless vector fields $A_\mu^a$ 
\begin{equation}
\mathcal{L}=-\frac{1}{4}\mathcal{G}_{ab}F^{a\mu\nu}F^b_{\mu\nu}
\end{equation}
with $F^a_{\mu\nu} = \partial_\mu A^a_\nu -\partial_\nu A^a_\mu$ and $\mathcal{G}_{ab}$ a metric in the field space. The isometry group of this metric leads to the presence of global symmetries that, through Noether theorem, gives rise to a set of conserved currents. Then, when the interactions for the fields are introduced as consistent couplings to the currents, again the resulting interactions are given by the Yang-Mills Lagrangian
\begin{equation}
\mathcal{L}=-\frac{1}{4}\mathcal{G}_{ab}\mathcal{F}^{a\mu\nu}\mathcal{F}^b_{\mu\nu}
\end{equation}
with the non-abelian field strength
\begin{equation}
\mathcal{F}^a_{\mu\nu} =F^a_{\mu\nu} + g f^{abc}A^b_\mu A^c_\nu
\label{eq:defF}
\end{equation}
where $g$ is the coupling constant of the non-abelian field and $f^{abc}$ are the structure constants of the Lie algebra of the isometry group of $\mathcal{G}_{ab}$ whose generators $T_a$ then satisfy $[T_a, T_b]=i f_{ab}{}^cT_c$ and can be normalized so that $\Tr (T_ a T_b)=\mathcal{G}_{ab}$, i.e., $\mathcal{G}_{ab}$ is nothing but the corresponding Killing metric of the group. The vector fields then take values in the Lie algebra of the group so that under an isometry transformation with parameters $\theta^a$ the vectors transform in the adjoint representation $A_\mu^a \to A_\mu^a + f_{bc}{}^a\theta^b A^c_\mu-\partial_\mu \theta^a/g$, i.e., as it corresponds to a connection. One can then introduce the covariant derivative  $D_\mu\equiv \partial_\mu \Id-ig A^a_\mu T_a$, whose commutator gives, as usual, the curvature: $[D_\mu,D_\nu]=-ig\mathcal{F}^a_{\mu\nu}T_a$. The field strength transforms covariantly\footnote{As opposed to the abelian case where the field strength is gauge invariant.} and, thus, the above Lagrangian is gauge invariant. Adding a mass term for the vector fields breaks the non-abelian gauge symmetry. This can be done either by adding a hard mass term to the Lagrangian or through a Higgs mechanism so that the gauge symmetry is spontaneously broken and it is non-linearly realised. Either way, the resulting Lagrangian will read
\begin{equation}
\mathcal{L}=-\frac{1}{4}\mathcal{G}_{ab}\mathcal{F}^{a\mu\nu}\mathcal{F}^b_{\mu\nu}-\frac12M_{ab} A_\mu^a A^{b\mu} .
\end{equation}
with $M_{ab}$ the mass matrix. Although the gauge symmetry is broken by the mass term, the original global symmetry can remain if, for instance, $M_{ab}\propto \mathcal{G}_{ab}$. Our main goal in this work is to construct the generalization of the massive non-abelian vector field to include derivative self-interactions. For the construction we will follow closely the approach applied in \cite{JbJLH}. For the sake of concreteness, we will restrict our analysis to the case of an internal rotational group for the vector fields, which can be viewed as a descendent of an original $SU(2)$ gauge symmetry. In that case, we have that the Killing metric is the Euclidian metric $\delta_{ab}$ (so lowering and raising group indices will be innocuous operations ) and the structure constants are given by the completely antisymmetric Levi-Civita symbol $\epsilon_{abc}$. Furthermore, we will assume that the global symmetry remains so that the number of possible interactions is substantially reduced. Since for $SO(3)$ the adjoint and the fundamental representations are equivalent, we will no make any distinction in the following.
\section{Systematical construction}
\label{Sec:SystConst}
In this section we will systematically construct the healthy derivate self-interactions for a set of vector fields with an internal global $SO(3)$ symmetry, as explained above. The procedure that we will follow is then based in the usual construction making use of the antisymmetry of the Levi-Civita tensor $\epsilon^{\mu\nu\alpha\beta}$ and that has been extensively exploited in the literature to construct healthy interactions. In particular, it was used in \cite{Deffayet:2010zh} to generalize the Galileon interactions to the case of arbitrary $p$-forms. For a set of interacting 1-forms (resembling the case under study here) it is possible to write Galileon interactions while retaining a non-abelian gauge symmetry. However, the first dimension where they are non-trivial is $D=5$ and, since our analysis will be performed in $D=4$, this result will prevent us from finding interactions respecting the gauge symmetry. For that reason, we will only consider terms with up to one derivative per field. Furthermore, we will impose a global rotational symmetry  by appropriately contracting the internal indices with $\delta_{ab}$ and  $\epsilon_{abc}$. It will be convenient to use $F_{\mu\nu}^a$ as defined above and $S_{\mu\nu}^a\equiv \partial_\mu A_\nu^a+\partial_\nu A_\mu^a$ to construct the healthy derivative interactions. This will allow us to clearly distinguish between the more traditional interactions where the derivatives of the vector fields only enter through $F_{\mu\nu}^a$ and the novel multi-Galileon inspired interactions that will explicitly contain $S_{\mu\nu}^a$. In analogy with the case of one single vector field, all the former interactions will be included in the term
\begin{equation}
\mathcal{L}_2=f_2(A_\mu^a, F^a_{\mu\nu})
\end{equation}
and our study will focus on the possible terms that explicitly contain $S_{\mu\nu}^a$. 

We will now proceed similarly to the systematic construction in \cite{JbJLH} order by order in derivatives of $A_\mu^a$ contracted with antisymmetric Levi-Civita tensors and vector fields, i.e., we will look for terms with the schematic form
\begin{eqnarray}
\mathcal{L}=f \epsilon^{\mu\nu\alpha\beta}\epsilon^{\rho\sigma\gamma\delta}\partial_\mu A^a_\rho\cdots  A^b_\nu A^c_\sigma\cdots  
\end{eqnarray}
where $f$ is a scalar function of the vector fields. The free Lorentz indices will be appropriately contracted with the spacetime metric and the internal indices will be contracted either with the group metric $\delta_{ab}$ or with $\epsilon_{abc}$. Notice that because of the structure of the interactions, no parity violating terms will be generated.

\subsection{Order $\partial A$}
\label{Subsec:dA}
At first order in derivatives we have the object $\partial_\mu A_\nu^a$ so that in order to have a Lorentz scalar and respect the global $SO(3)$ symmetry we need to add an even number of vector fields in the schematic form\footnote{We introduce the notation $[\cdots]$ to denote a scalar quantity with respect to Lorentz and internal indices of the object inside the brackets.} $\LL_3^{2nA}\sim[\epsilon\epsilon\partial A A^{2n}]$. At this order it is not possible to construct a term with $n=0$ due to the floating internal index, hence $\mathcal{L}_3^{(0A)}=0$. This means that this case is different from the single vector field case where a term with no $A$'s is possible (corresponding to the generalization of the cubic vector Galileon interaction $f_3(A^2)[S]$). This is simply a consequence of imposing the global symmetry and, thus, at this order it is not possible to write terms that preserve the symmetry. The first non-trivial terms require $n=1$ and can be written as
\begin{eqnarray}\label{Testeq_L32A}
\mathcal{L}_3^{(2A)}&=&f_{3,1} \epsilon^{\mu\nu\alpha\beta}\epsilon^{\rho\sigma}{}_{\alpha\beta}\partial_\mu A^a_\rho A^b_\nu A^c_\sigma  \epsilon_{abc} \nonumber\\
&+&f_{3,2} \epsilon^{\mu\nu\alpha\beta}\epsilon^{\rho\sigma}{}_{\alpha\beta}\partial_\mu A^a_\nu A^b_\rho A^c_\sigma \epsilon_{abc}
\end{eqnarray}
where $f_{3,i}$ are scalar functions of the vector fields. Let us mention here that, at this order, we could also have a parity violating term as $\epsilon^{\gamma\mu\alpha\rho}\partial_\mu A^a_\alpha A^c_\gamma A^b_\rho  \epsilon_{abc} $. However, as mentioned earlier, our procedure preserves parity and, thus, those terms cannot be directly generated. Since we restrict our analysis to parity preserving interactions, we will not consider these type of terms from now on, although they could be constructed in an analogous systematic way in terms of only one spacetime Levi-Civita tensor. Another possibility would be the term $\epsilon^{bc}{}_a \partial_\mu A^a_\alpha A^\mu_b A^\alpha_c$, but this gives the same contribution as those already included in the above expression. 
We could also consider terms where three of the spacetime $\epsilon$'s indices are contracted as $ \epsilon^{\mu\nu\alpha\beta}\epsilon^{\rho}{}_{\nu\alpha\beta}\partial_\mu A^a_\rho A^{b\sigma} A^c_\sigma  \epsilon_{abc}$ but this is identically zero due to the antisymmetry of the $ \epsilon_{abc} $ tensor. In fact, a closer inspection of the two terms in (\ref{Testeq_L32A}) reveals that only the antisymmetric part of $\partial_\mu A^a_\nu$ contributes to $\mathcal{L}_3^{(2A)}$  so that both terms can be written as $F^{a\mu\nu}A^b_\mu A^c_\nu\epsilon_{abc}$ and, therefore, they are included in $\mathcal{L}_2$. This is due to the fact that the group indices are contracted with the antisymmetric $ \epsilon_{abc} $ tensor. After all these considerations, we conclude that all the interactions at this order can be summarized as
\begin{eqnarray}
\mathcal{L}_3^{(2A)}=f_3F_{\mu\nu}^a A^\mu_b A^\nu_c \epsilon_a{}^{bc} \; \subset \; \mathcal{L}_2.
\end{eqnarray}
At this stage, we find that it is not possible to construct terms of the type $[SAA]$ that respect the required global symmetry, since the antisymmetric index structure of the group indices does not allow it. This is again an important difference with respect to the single vector field case, where this type of interactions are possible. We will comment more on this below. In order to have non-trivial interactions with one derivative that are not included in $\mathcal{L}_2$ we need to go to the next order with $n=2$, i.e, we need to contract with four vector fields in the schematic form $[\epsilon \epsilon \partial A A^4]$. The corresponding standard contractions of the indices
\begin{eqnarray}
\mathcal{L}_3^{(4A)}&\supset&g_{3,1} \epsilon^{\mu\nu\alpha\beta}\epsilon^{\rho\sigma\delta}{}_{\beta}\partial_\mu A^a_\rho A^b_\nu A^c_\sigma  A^d_\alpha A^e_\delta  \epsilon_{abc} \delta_{de} \nonumber\\
&+&g_{3,2} \epsilon^{\mu\nu\alpha\beta}\epsilon^{\rho\sigma\delta}{}_{\beta}\partial_\mu A^a_\nu A^b_\alpha A^c_\rho  A^d_\sigma A^e_\delta  \epsilon_{abc} \delta_{de}
\end{eqnarray}
are again such that they correspond to terms of the form $[FA^4]$ belonging to $\mathcal{L}_2$. This is also the case if the internal indices were contracted differently, for instance with $\epsilon_{bde} \delta_{ac}$. Irrespectively of the contraction of the internal indices, if all of the Lorentz indices are contracted with the two Levi-Civita tensors, then at this order we can only construct terms of the type $[FA^4]$. Additional types of new interactions will arise only if the contractions of the indices do not select the  antisymmetric part of $\partial A$ as in the previous terms, but they allow for the symmetric part of $\partial A$ generating terms of the form $[SA^4]$. In order for this to be the case we have to contract the Lorentz indices of two vector fields between themselves
\be
\mathcal{L}_3^{(4A)}\supset g_{3,3} \epsilon^{\mu\nu\alpha\beta}\epsilon^{\rho\sigma}{}_{\alpha\beta}\partial_\mu A^a_\rho A^b_\nu A^c_\sigma  A^d_\delta A^{e \delta}  \epsilon_{abe} \delta_{cd}.
\ee
Leaving aside terms of the form $[FA^4]$ that are already included in $\LL_2$, the above interaction gives rise to the following new term
\begin{eqnarray}
\mathcal{L}_3^{(4A)} \supset g_{3,3} S^{a\mu\nu} A_\mu^b A_\nu^d A_\alpha^c A^{e\alpha} \delta_{de} \epsilon_{abc}.
\end{eqnarray}
Moreover, by symmetry this is in fact the only non-trivial term at this order that is not included in $\mathcal{L}_2$ so that we have that $\mathcal{L}_3^{(4A)}$ leads to
\begin{eqnarray}
\mathcal{L}_3^{(4A)} \to \mathcal{L}_2+g_{3} S^{a\mu\nu} A_\mu^b A_\nu^d A_\alpha^c A^{e \alpha} \delta_{de} \epsilon_{abc},
\end{eqnarray}
 with $g_3$ again a scalar function of the vector fields. 
 
 Finally, the next order with $n=3$ includes contractions with six vector fields $[\epsilon \epsilon \partial A A^6]$ and this saturates the indices of the spacetime Levi-Civita tensor and, thus, our series will stop at this order.  If we contract the indices of all the vector fields with the Levi-Civita tensors $ \epsilon^{\mu\nu\alpha\beta}\epsilon^{\rho\sigma\delta\gamma}\partial_\mu A^a_\rho A_\nu^b A_\sigma^c A_\alpha^d A^{e}_{\delta} A^{f\beta} A_\gamma^g$, then this order will produce either vanishing interactions or again interactions of the form $[FA^6]$ that are included in $\mathcal{L}_2$. In order to construct interactions of the form $[SA^6]$ we again need to contract the Lorentz indices of the vectors with no derivatives among themselves, yielding the following three possibilities
 \begin{align}
&\mathcal{L}_3^{(6A)} =  \epsilon^{\mu\nu\alpha\beta}\epsilon^{\rho\sigma}{}_{\alpha\beta} \partial_\mu A^a_\rho A_\nu^b A_\sigma^c A_\lambda^d A^{e\lambda} A^{f\delta} A_\delta^g\nonumber\\
&\times\Big(h_{3,1}\epsilon_{abd}\delta_{cf}\delta_{eg}+h_{3,2}\epsilon_{bdf}\delta_{ae}\delta_{cg}+h_{3,3}\epsilon_{abe}\delta_{cd}\delta_{fg}\Big)
\end{align}
with $h_{3,i}$ being again scalar functions of the vector fields $A_\mu^a$. We could also contract the additional four vector fields in the form $\epsilon^{\mu\nu\alpha\beta}\epsilon^{\rho\sigma}{}_{\alpha\beta} \partial_\mu A^a_\rho A_\nu^b A_\sigma^c A_\lambda^d A^{e}_\delta A^{f\lambda} A^{g\delta}$ but this is already included in the above interactions. Apart from the terms $[FA^6]$, these interactions will result in the new terms
\begin{align}
&\mathcal{L}_3^{(6A)} = S^{a\mu\nu} A_\mu^b A_\nu^c A_\alpha^d A^{e\alpha} A^{f\beta} A_\beta^g \nonumber\\
&\times\Big(h_{3,1}\epsilon_{abd}\delta_{cf}\delta_{eg}+h_{3,2}\epsilon_{bdf}\delta_{ae}\delta_{cg}+h_{3,3}\epsilon_{abe}\delta_{cd}\delta_{fg}\Big).
\end{align}
It is also worth mentioning that some of the non-trivial contractions like e.g. $S^{a\mu\nu} A_\mu^b A_\nu^c A_\alpha^d A^{e\alpha} A^{f\beta} A_\beta^g \epsilon_{abd}\delta_{ce}\delta_{fg}$ are already contained in $\LL_3^{(4A)}$.

Since at this order there are no more free indices of the Levi-Civita tensors to be contracted with additional vector fields the series ends here and we need to go to the next order in derivatives to construct new interactions. One might wonder, if one could simply add additional pairs of vector fields whose Lorentz indices are contracted together but not their internal indices, such that one could generate terms of the form $ \epsilon^{\mu\nu\alpha\beta}\epsilon^{\rho\sigma}{}_{\alpha\beta} \partial_\mu A^a_\rho A_\nu^b A_\sigma^c A_\lambda^d A^{e\lambda} A^{f\delta} A_\delta^g A^{h\kappa} A^i_\kappa$ resulting in interactions of the form $[SA^8]$. The answer is obviously affirmative and, in fact, we can extend the obtained interactions by simply replacing $\delta_{ab}\to\delta_{ab}+ A_{a\alpha} A_b^\alpha$ or $\eta_{\mu\nu}\to\eta_{\mu\nu}+A^a_\mu A_{a\nu}$, possibly with some additional scalar functions in front of each term. This procedure to generate higher order interactions will be valid for all the interactions obtained below and, in fact, we will see that some interactions at a given order can be obtained from previous orders by applying this procedure.

This order was not too delicate as it contained only one derivative and, thus, it is not possible to excite an additional polarization in any case that could jeopardize the stability of the theory. However, it already presented some interesting features like the impossibility of building certain interactions while preserving the global symmetry and, in addition, it allowed to present in a detailed way the general reasoning that will be used in the more interesting higher order interactions treated in the rest of this section.

\subsection{Order $(\partial A)^2$}
Now we will consider interactions with two derivatives of the schematic form $[\epsilon\epsilon\partial A\partial A A^{2n}]$. Unlike the previous subsection with only one derivative, now we can have interactions for $n=0$ respecting the global symmetry. These terms resemble those of the single field case and can be written as
\begin{eqnarray}
\mathcal{L}_4^{(0A)}&=& f_{4,1}\epsilon^{\mu\nu\alpha\beta}\epsilon^{\rho\sigma}{}_{\alpha\beta} \partial_\mu A^a_\rho \partial_\nu A^b_\sigma \delta_{ab}  \nonumber\\
&+&f_{4,2}\epsilon^{\mu\nu\alpha\beta}\epsilon^{\rho\sigma}{}_{\alpha\beta} \partial_\mu A^a_\nu \partial_\rho A^b_\sigma \delta_{ab} .  
\end{eqnarray}
The second term corresponds to the contraction of two field strength tensors $F^{\mu\nu}_a F_{\mu\nu}^a$ so that it can be again simply included in $\mathcal{L}_2$. The first term is the direct extension of the interaction that we are familiar with from the generalized Proca field $[S^2]-[S]^2$ to the case of several vector fields. Thus, this order gives rise to
\begin{eqnarray}\label{LagL4nonAproca}
\mathcal{L}_4^{(0A)} \to   \mathcal{L}_2 + f_4 \Big(S_{\mu\nu}^aS^{b\mu\nu}-S_\mu^{a\mu}S_{\nu }^{b\nu}\Big)\delta_{ab}.
\end{eqnarray}
We see that at this order with two derivatives it is possible to directly extend the generalized Proca interaction to the present case and preserving the global symmetry, unlike the previous order with only one derivative where this was not possible.

Now we can proceed to the next order with two vector fields corresponding to $n=1$. The relevant terms at this order are
\begin{align}
\mathcal{L}_4^{(2A)}\supset&\epsilon^{\mu\nu\gamma\alpha}\epsilon^{\rho\sigma\delta}{}_{\alpha} \partial_\mu A^a_\rho \partial_\nu A^b_\sigma A_\gamma^cA_\delta^d\big(g_{4,1}
\delta_{ab}\delta_{cd}+g_{4,2}
\delta_{ac}\delta_{bd}\big)\nonumber\\
&g_{4,3}\epsilon^{\mu\nu\gamma\alpha}\epsilon^{\rho\sigma\delta}{}_{\alpha} \partial_\mu A^a_\nu \partial_\gamma A^b_\rho A_\sigma^cA_\delta^d\delta_{ac}\delta_{bd},
\label{L42Aepsilon}
\end{align}
with $g_{4,i}$ scalar functions of $A_\mu^a$. We have left out other possible contractions like $\epsilon^{\mu\nu\gamma\alpha}\epsilon^{\rho\sigma\delta}{}_{\alpha} \partial_\mu A^a_\nu \partial_\rho A^b_\sigma A_\gamma^cA_\delta^d$ since they give rise to interactions of the form $[FFAA]$ or interactions already included in the previous orders. The terms in the first line result in contributions of the form $[SSAA]$. More explicitly,  up to terms $[FFAA]$ already included in $\LL_2$, we find the following contributions from the first line in (\ref{L42Aepsilon}):
\begin{align}\label{candL42A}
\mathcal{L}_4^{(2A)}\supset & g_{4,1}\Big[ \Big(S_{\mu\nu}^aS^{b\mu\nu}-S_\mu^{a\mu}S_{\nu }^{b\nu}\Big)\delta_{ab}A^2\nonumber\\
&+2\Big(S_\lambda^{a\lambda}S^{a\mu\nu}-S^{a\mu}{}_\lambda S^{a\lambda\nu}\Big)A^b_\mu A^b_\nu\Big]\nonumber\\
+&g_{4,2}\Big[\Big(S_{\mu\nu}^aS^{b\mu\nu}-S_\mu^{a\mu}S_{\nu }^{b\nu}\Big)A^{a\lambda}A^b_\lambda\nonumber\\
&+2\Big(S_\lambda^{a\lambda}S^{b\mu\nu}-\frac12S^{a(\mu}{}_\lambda S^{b\nu)\lambda}\Big)A^a_\mu A^b_\nu\Big].
\end{align}
The first line of this expression can be absorbed into $\LL_{4}^{(0A)}$ in (\ref{LagL4nonAproca}) via a redefinition of $f_4$, while the remaining new interactions genuinely belong to $\LL_4^{(2A)}$. We should notice that the term proportional to $g_{4,2}$ actually gives two different terms with a fixed relative coefficient (a factor of 2). This is accidental and, in fact, both terms can come in with different functions. The degeneracy can be broken in a very simple way by using the discussion at the end of \ref{Subsec:dA}. We can simply replace $\delta_{ab}\to g A_{a\lambda} A_{b}^\lambda$ with $g$ a scalar function in (\ref{LagL4nonAproca}) and, thus, the degeneracy will be broken. Following the systematic procedure, the breaking of the degeneracy can be seen to occur with the term
\begin{align}
\mathcal{L}_4^{(2A)}\supset& \;2g_{4,4}\epsilon^{\mu\nu\alpha\beta}\epsilon^{\rho\sigma}{}_{\alpha\beta} \partial_\mu A^a_\rho \partial_\nu A^b_\sigma A_\gamma^cA^{d\gamma}\delta_{ac}\delta_{bd}\nonumber \\
=&g_{4,4}\Big(S_{\mu\nu}^aS^{b\mu\nu}-S_\mu^{a\mu}S_{\nu }^{b\nu}\Big)A^{a\lambda}A^b_\lambda\, .
\end{align}
Thus, the presence of this interaction will detune the relative coefficients of the terms proportional to $g_{4,2}$ in equation (\ref{candL42A}). Furthermore, the other similar contraction $\epsilon^{\mu\nu\alpha\beta}\epsilon^{\rho\sigma}{}_{\alpha\beta} \partial_\mu A^a_\rho \partial_\nu A^b_\sigma A_\gamma^cA^{d\gamma}
\delta_{ab}\delta_{cd}$ will again give a contribution to $\LL_{4}^{(0A)}$ in equation (\ref{LagL4nonAproca}).

On the other hand, the interaction proportional to $g_{4,3}$ in (\ref{L42Aepsilon}) gives rise to an additional term of the form $[FSAA]$ that can be written as 
\begin{eqnarray}
\mathcal{L}_4^{(2A)}\supset  g_{4,3}\epsilon^{\alpha\beta\gamma\delta}\tilde{F}^a_{\alpha\lambda} S^{b\lambda}{}_\beta A^a_\gamma A^b_\delta.
\end{eqnarray}
Notice that this term cannot exist for the single vector field case and, therefore, it is genuine of the multi-Proca theory. The interactions at this order can be summarized as
\begin{align}
\mathcal{L}_4^{(2A)} \to&   \mathcal{L}_2 +\LL_4^{(0A)}
+g_{4,1}\Big(S_\lambda^{a\lambda}S^{a\mu\nu}-S^{a\mu}{}_\lambda S^{a\lambda\nu}\Big)A^b_\mu A^b_\nu\nonumber\\
&+g_{4,2}\Big(S_\lambda^{a\lambda}S^{b\mu\nu}-\frac12S^{a(\mu}{}_\lambda S^{b\nu )\lambda}\Big)A^a_\mu A^b_\nu \nonumber\\
&+g_{4,3}\epsilon^{\alpha\beta\gamma\delta}\tilde{F}^a_{\alpha\lambda} S^{b\lambda}{}_\beta A^a_\gamma A^b_\delta \nonumber\\
&+g_{4,4}\Big(S_{\mu\nu}^aS^{b\mu\nu}-S_\mu^{a\mu}S_{\nu }^{b\nu}\Big)A^{a\lambda}A^b_\lambda.
\end{align}
The next order corresponds to $n=2$ so that we will have interactions of the form $[\epsilon\epsilon (\partial A)^2 A^4]$. Including the additional vector fields gives the possible contractions
\begin{align}
 \mathcal{L}_4^{(4A)} &\supset h_{4,1} \epsilon^{\mu\nu\gamma\alpha}\epsilon^{\rho\sigma\delta\beta} \partial_\mu A^a_\rho \partial_\nu A^c_\sigma A_\gamma^bA_\delta^d A_\alpha^e A_\beta^f \epsilon_{abe}\epsilon_{cdf} \nonumber\\
    &+h_{4,2}\epsilon^{\mu\nu\gamma\alpha}\epsilon^{\rho\sigma\delta\beta} \partial_\mu A^a_\nu \partial_\gamma A^c_\rho A_\alpha^bA_\sigma^d A_\delta^e A_\beta^f \epsilon_{abe}\epsilon_{cdf} .
\end{align}
The first line of this expression gives rise to interactions of the form $[SSA^4]$ and $[FFA^4]$. Omitting as usual the terms $[FFA^4]$ that will contribute to $\LL_2$, the new interactions from the first line yield couplings of two $S$'s and four $A$'s contracted in the appropriate ways to avoid extra propagating polarizations. We omit here their long expression in terms of $S$, but we will give a compact expression for it in \ref{sec:GenForm}. In any case, the exact coefficients can be easily computed from the above contraction with the Levi-Civita tensors. From the second line, on the other hand, we obtain interactions of the mixed form $[FSA^4]$. We have not considered the interaction $\epsilon^{\mu\nu\gamma\alpha}\epsilon^{\rho\sigma\delta\beta} \partial_\mu A^a_\nu \partial_\rho A^c_\sigma A_\gamma^bA_\delta^d A_\alpha^e A_\beta^f$ since this corresponds to a term of the form  $[FFA^4]$ as well, which is already part of $\mathcal{L}_2$. We could also construct $\partial_\mu A_\alpha^a\partial_\nu A_\beta^b A^\mu_c A^\alpha_d \epsilon^{cd}{}_a A^\nu_e A^\beta_f \epsilon^{ef}{}_b $, but this is already included in the above interactions. Furthermore, instead of contracting the internal indices with $\epsilon_{abe}\epsilon_{cdf}$, we could have contracted them with $\delta_{ab}\delta_{cd}\delta_{ef}$ and permutations of them. However they correspond to the previous order with again the replacement $\delta_{ab}\to A_{a\lambda}A_b^\lambda$. In a similar way as in the previous order, we could also contract the indices of the vector fields among themselves 
\begin{align}
 \mathcal{L}_4^{(4A)} &\supset h_{4,3} \epsilon^{\mu\nu\gamma\alpha}\epsilon^{\rho\sigma\delta}{}_\alpha \partial_\mu A^a_\rho \partial_\nu A^c_\sigma A_\gamma^bA_\delta^d A^{e\beta} A_\beta^f \epsilon_{abe}\epsilon_{cdf} \nonumber\\
&+h_{4,4}  \epsilon^{\mu\nu\gamma\alpha}\epsilon^{\rho\sigma\delta}{}_\alpha \partial_\mu A^a_\nu \partial_\gamma A^c_\rho A_\sigma^bA_\delta^d A^{e\beta} A_\beta^f \epsilon_{abe}\epsilon_{cdf} 
\end{align}
They also give contributions in form of $[SSA^4]$ and $[FSA^4]$ respectively, apart from the $[FFA^4]$ terms. With four additional vector fields at this order we already used all the Lorentz indices of the two Levi-Civita tensors, and hence our construction stops here. Let us mention however that we could generate higher order terms by the procedure of replacing $\delta_{ab}\to A_{a\lambda} A_b^\lambda$, resulting in terms like
$\partial_\mu A_\nu^a\partial_\alpha A_\beta^b A_\rho^c \epsilon_{abc} A^\mu_e A^\nu_f A^\alpha_g \epsilon^{efg} A^\beta_m A^\rho_n \delta^{mn}$. This term goes beyond the orders comprised within our systematical construction using the Levi-Civita tensors.

\subsection{Order $(\partial A)^3$}
The interactions at this order in derivatives will have the schematic form $[\epsilon\epsilon\partial A\partial A\partial A A^{2n}]$. Starting with the $n=0$ terms, we can construct
\begin{eqnarray}
\mathcal{L}_5^{(0A)}&=& f_{5,1}\epsilon^{\mu\nu\alpha\beta}\epsilon^{\rho\sigma\delta}{}_{\beta} \partial_\mu A^a_\rho \partial_\nu A^b_\sigma \partial_\alpha A^c_\delta \epsilon_{abc}  \nonumber\\
&+& f_{5,2}\epsilon^{\mu\nu\alpha\beta}\epsilon^{\rho\sigma\delta}{}_{\beta} \partial_\mu A^a_\nu \partial_\rho A^b_\sigma \partial_\alpha A^c_\delta \epsilon_{abc}.
\end{eqnarray}
At this order, because of the contraction with $\epsilon_{abc}$ the interactions of the first line vanish identically. This means that the terms cubic in $S$ vanish. Thus, the direct extension of the single vector field interactions $[S]^3-3[S][S^2] +2[S^3]$ at that order to the case of several vector fields with the global symmetry is again not possible. Similarly, the extension of the interaction $\Ft \Ft S$ for the generalized Proca field that should arise from the second line is not allowed by the global symmetry. The only surviving term from the second line is of the form $FFF\epsilon$, which is already part of $\mathcal{L}_2$. Hence, we do not have any new interactions at this order
\begin{eqnarray}
\mathcal{L}_5^{(0A)}\subset  \mathcal{L}_2.
\end{eqnarray}
We can then go to the next order with $n=1$ to construct the terms of the form $[(\partial A)^3A^2]$. The possible contractions are
\begin{eqnarray}
\mathcal{L}_5&=& 
 g_{5,1}\epsilon^{\mu\nu\alpha\beta}\epsilon^{\rho\sigma\delta\gamma} \partial_\mu A^a_\rho \partial_\nu A^b_\sigma \partial_\alpha A^c_\delta A^d_\beta A^e_\gamma \epsilon_{dea} \delta_{bc}\nonumber\\
&+&  \epsilon^{\mu\nu\alpha\beta}\epsilon^{\rho\sigma\delta\gamma}\partial_\mu A^a_\nu \partial_\rho A^b_\sigma \partial_\alpha A^c_\delta A^d_\beta A^e_\gamma \nonumber\\
 &\times&(g_{5,2} \epsilon_{dea} \delta_{bc}+g_{5,3}\epsilon_{acd} \delta_{be}+g_{5,4}\epsilon_{abd} \delta_{ce}) .
\end{eqnarray}
From the first line of the above expression we obtain interactions of the schematic form $[FSSAA]$ with the specific contraction of the Lorentz and internal indices dictated by the Levi-Civita tensors. Their exact expression is cumbersome and should be taken from the above contraction. On the other hand, from the second line we obtain terms of the form $[FFSAA]$. We have omitted the contribution $\epsilon^{\mu\nu\alpha\beta}\epsilon^{\rho\sigma\delta\gamma}\partial_\mu A^a_\nu \partial_\rho A^b_\sigma \partial_\alpha A^c_\beta A^d_\delta A^e_\gamma$ since it  only contributes terms $[F^3A^2]\subset \mathcal{L}_2$. In a similar way as in the previous order, we could also try to  contract two of the Lorentz indices of the vectors without derivatives between themselves, i.e. considering terms of the form $\epsilon^{\mu\nu\alpha\beta}\epsilon^{\rho\sigma\delta}{}_\beta \partial_\mu A^a_\rho \partial_\nu A^b_\sigma \partial_\alpha A^{c}_\delta A^{d\gamma} A^e_\gamma$. However, it turns out that this contraction vanishes identically, hence we can not construct terms of the form $[SSSAA]$.

\subsection{Order $(\partial A)^4$}
Finally, we will consider interactions with 4 derivatives and the contractions with the Levi-Civita tensors are then already saturated with the terms with derivatives, i.e., there are no free indices to be contracted with additional vector fields without derivatives. The new interactions that we obtain at this order are
\begin{eqnarray}\label{eq_L60A}
\mathcal{L}_6^{(0A)}&=&\epsilon^{\mu\nu\alpha\beta}\epsilon^{\rho\sigma\delta\gamma} \partial_\mu A^a_\rho \partial_\nu A^b_\sigma \partial_\alpha A^c_\delta \partial_\beta A^d_\gamma \nonumber\\
&&\times(f_{6,1}\delta_{ab}\delta_{cd}+f_{6,2}\delta_{ac}\delta_{bd}) \nonumber\\
&+&\epsilon^{\mu\nu\alpha\beta}\epsilon^{\rho\sigma\delta\gamma} \partial_\mu A^a_\nu \partial_\rho A^b_\sigma \partial_\alpha A^c_\delta \partial_\beta A^d_\gamma  \nonumber\\
&&\times (f_{6,3}\delta_{ab}\delta_{cd}+f_{6,4}\delta_{ac}\delta_{bd}) 
\end{eqnarray}
The first line of this equation gives three types of interactions of the schematic form $[F^4]$, $[S^4]$ and $[F^2S^2]$ respectively, whereas the second line gives contributions in form of $[F^4]$ and $[F^2S^2]$. We can again absorb the purely $F$ terms into  $\mathcal{L}_2$ and neglect them here. This is also the reason why we omitted the term $\epsilon^{\mu\nu\alpha\beta}\epsilon^{\rho\sigma\delta\gamma} \partial_\mu A^a_\nu \partial_\rho A^b_\sigma \partial_\alpha A^c_\beta \partial_\delta A^d_\gamma$. 
The interactions quartic in $S$ are just total derivatives and the $[F^2S^2]$ interactions have the two possible contractions of the internal indices $\Ft^{a\mu\nu}\Ft^{\alpha\beta}_a S^b_{\mu\alpha} S_{b\nu\beta}$ and $\Ft^{a\mu\nu}\Ft^{b\alpha\beta} S_{a\mu\alpha} S_{b\nu\beta}$, i.e. the terms that we are familiar with from the single massive vector field. 

\section{General form of the Lagrangian and covariantization}
\label{sec:GenForm}
In the previous section, we have systematically constructed the possible interactions for a set of vector fields order by order using the  Levi-Civita tensors. One can easily check that, in fact, all the different terms can be compactly written as
\begin{align}
\LL_2=&G_2(A_\mu^a, F_{\mu\nu}^a)\;,\nonumber\\
\LL_3=&\K_a^{\mu\nu}S^a_{\mu\nu}\;,\nonumber\\
\LL_4=&\K_{ab}^{\mu_1\mu_2\nu_1\nu_2} S^a_{\mu_1\nu_1}S^b_{\mu_2\nu_2}\nonumber\\
&+\M_{ab}^{\mu_1\mu_2\nu_1\nu_2} F^a_{\mu_1\nu_1}S^b_{\mu_2\nu_2}\;,\nonumber\\
\LL_5=&\M_{abc}^{\mu_1\mu_2\mu_3\nu_1\nu_2\nu_3}F^a_{\mu_1\nu_1}S^b_{\mu_2\nu_2} S^c_{\mu_3\nu_3} \nonumber\\
&+\N_{abc}^{\mu_1\mu_2\mu_3\nu_1\nu_2\nu_3}F^a_{\mu_1\nu_1}F^b_{\mu_2\nu_2} S^c_{\mu_3\nu_3} \nonumber\\
&+\hat{\N}_{abc}^{\mu_1\mu_2\mu_3\nu_1\nu_2\nu_3}F^a_{\mu_1\mu_2}F^b_{\nu_1\nu_2} S^c_{\mu_3\nu_3}\;, \nonumber\\
\LL_6=&\N_{abcd}^{\mu_1\mu_2\mu_3\mu_4\nu_1\nu_2\nu_3\nu_4}F^a_{\mu_1\mu_2}F^b_{\nu_1\nu_2} S^c_{\mu_3\nu_3} S^d_{\mu_4\nu_4}\nonumber\\
&+\hat{\N}_{abcd}^{\mu_1\mu_2\mu_3\mu_4\nu_1\nu_2\nu_3\nu_4}F^a_{\mu_1\nu_1}F^b_{\mu_2\nu_2} S^c_{\mu_3\nu_3} S^d_{\mu_4\nu_5}\;,
\label{Eq:FullLagrangian}
\end{align}
where $\K_{a\dots}^{\mu_1\dots\nu_1\dots}$, $\M_{a\dots}^{\mu_1\dots\nu_1\dots}$, $\N_{a\dots}^{\mu_1\dots\nu_1\dots}$ and $\hat{\N}_{a\dots}^{\mu_1\dots\nu_1\dots}$ are objects built out of $A_\mu^a$ (in addition to the spacetime and group metrics and their corresponding Levi-Civita tensors) such that they are completely antisymmetric in the indices $\mu_i$ and $\nu_i$ separately. Notice that the antisymmetry of these objects will also guarantee the absence of higher order equations for the transverse modes thanks to the Bianchi identities $\nabla_{[\alpha}F^a_{\mu\nu]}=0$. 
To better illustrate the general form of the interactions expressed in (\ref{Eq:FullLagrangian}) we will give the explicit expressions at lowest order in $A_\mu^a$ for a few non-trivial examples:
\begin{align}
&\K^ {\mu\nu}_a=\epsilon_{abc}\delta_{de}A^{b \mu} A^{d\nu} A^c_\lambda A^{e\lambda}+\cdots\\
&\K^{\mu_1\mu_2\nu_1\nu_2}_{ab}=\epsilon^{\mu_1\mu_2\mu_3\lambda}\epsilon^{\nu_1\nu_2\nu_3}{}_\lambda\Big(c_1\delta_{ab}\delta_{cd}+c_2\delta_{ac}\delta_{bd}\Big)\nonumber\\
&\hspace{1.7cm}\times\Big(d_1A^c_{\mu_3}A^d_{\nu_3} +d_2A^c_{\lambda}A^{d\lambda} \eta_{\mu_3\nu_3}\Big)+\cdots\\
&\M^{\mu_1\mu_2\nu_1\nu_2}_{ab}=\epsilon^{\mu_1\mu_2\mu_3\lambda}\epsilon^{\nu_1\nu_2\nu_3}{}_{\lambda} A_{a\mu_3}A_{b\nu_3}+\cdots
\end{align}
where $c_i$ and $d_i$ are some constants. The remaining ones can be straightforwardly found from our previous systematic construction in terms of the Levi-Civita tensors. Since we are considering four dimensions, the objects in $\LL_6$ must factorize into the product of two spacetime Levi-Civita tensors
\begin{align}
\N_{abcd}^{\mu_1\mu_2\mu_3\mu_4\nu_1\nu_2\nu_3\nu_4}=\N_{abcd} \epsilon^{\mu_1\mu_2\mu_3\mu_4}\epsilon^{\nu_1\nu_2\nu_3\nu_4}\\
\hat{\N}_{abcd}^{\mu_1\mu_2\mu_3\mu_4\nu_1\nu_2\nu_3\nu_4}=\hat{\N}_{abcd} \epsilon^{\mu_1\mu_2\mu_3\mu_4}\epsilon^{\nu_1\nu_2\nu_3\nu_4}
\end{align}
with $\N_{abcd}$ and $\hat{\N}_{abcd}$ Lorentz scalar functions with four internal indices built out of $A_\mu^a$. Moreover, the fact that there are no forms of rank higher than the dimension of the spacetime prevents the construction of higher order terms and, therefore, the series ends in $\LL_6$. This is another way of expressing that at 4th order in $\partial A$, the indices of the spacetime Levi-Civita tensors are saturated and, hence, the systematic construction ends at that order.

Written in the compact form given in (\ref{Eq:FullLagrangian}), it is simple to show that the interactions can be extended to curved spacetime by adding appropriate non-minimal couplings. The terms linear in $S^a_{\mu\nu}$ can be directly extended to curved spacetime without adding any non-minimal couplings. The reason is that those interactions are linear in the connection so the corresponding generalization to curved spacetime can only generate additional terms with first derivatives of the vector fields to the equations coming from the coupling of $A_\mu^a$ to the connection in $S^a_{\mu\nu}$. On the other hand, the terms that are quadratic in $S$ do require the introduction of non-minimal couplings to maintain the desired propagating degrees of freedom in arbitrary spacetimes. To compute the necessary counter-terms we will write all the terms quadratic in $S^a_{\mu\nu}$ in the compact form:
\be
\LL_{\Pm}=\Pm_{ab}^{\mu_1\mu_2\nu_1\nu_2}S^a_{\mu_1\nu_1} S^b_{\mu_2\nu_2}
\ee
where $\Pm_{ab}^{\mu_1\mu_2\nu_1\nu_2}$ receives contributions from all the terms in (\ref{Eq:FullLagrangian}) that are quadratic in $S^a_{\mu\nu}$. Notice that $\Pm$ depends on $A_\mu^a$ and $F^a_{\mu\nu}$. This term will contribute dangerous terms to the energy-momentum tensor coming from the variation of the connection given by
\be
\delta\LL_{\Pm}\supset\Pm_{ab}^{\mu_1\mu_2\nu_1\nu_2}S^a_{\mu_1\nu_1} \delta\Gamma^\lambda_{\mu_2\nu_ 2} A^b_{\lambda}
\label{varLSqu}
\ee
which, after integration by parts, will make $A_0$ dynamical. Thus, to avoid exciting additional undesired polarizations in non-trivial spacetime background, we need to add a counter term that we will write as
\be
\LL_{\Qm}=R_{\mu_1\mu_2\nu_1\nu_2}\Qm^{\mu_1\mu_2\nu_1\nu_2}.
\ee
where $\Qm^{\mu_1\mu_2\nu_1\nu_2}$ is an object with the same symmetries as the Riemann tensor and which can be tuned to cancel the contributions from (\ref{varLSqu}). When summing both terms $\LL=\LL_{\Pm}+\LL{\Qm}$ we will have
\be
\delta \LL=\Big[\partial_\beta\Qm^{\beta (\mu\nu)}{}_\alpha+A^a_\alpha\Pm^{\beta(\mu\nu)\gamma}_{(ab)}S^b_{\beta\gamma}\Big]\delta\Gamma^\alpha_{\mu\nu}+\cdots
\ee
where $\cdots$ denotes additional non-dangerous terms. We will avoid these dangerous terms if the bracket vanishes identically. We can further simplify this expression if we expand
\be
\partial_\beta\Qm^{\beta (\mu\nu)}{}_\alpha=
\frac{\partial \Qm^{\beta (\mu\nu)}{}_\alpha}{\partial A^b_\gamma}\partial_\beta A^b_\gamma
+\frac{\partial \Qm^{\beta (\mu\nu)}{}_\alpha}{\partial F^b_{\gamma\delta}}\partial_\beta F^b_{\gamma\delta}.
\ee
Thus, in order to avoid additional propagating polarizations for the vector field we need to impose
\begin{align}
\left[\frac12\frac{\partial \Qm^{\beta (\mu\nu)}{}_\alpha}{\partial A^b_\gamma}+A^a_\alpha\Pm^{\beta(\mu\nu)\gamma}_{(ab)}\right]S^b_{\beta \gamma}+\frac{\partial \Qm^{\beta (\mu\nu)}{}_\alpha}{\partial F^b_{\gamma\delta}}\partial_\beta F^b_{\gamma\delta}=0.
\end{align}
This condition is identically fulfilled if we impose the relation
\begin{align}
\frac12\frac{\partial \Qm^{\beta (\mu\nu)}{}_\alpha}{\partial A^b_\gamma}+A^a_\alpha\Pm^{\beta(\mu\nu)\gamma}_{(ab)}+(\beta\leftrightarrow\gamma)=0
\end{align}
and $\frac{\partial \Qm^{\beta (\mu\nu)}{}_\alpha}{\partial F^b_{\gamma\delta}}$
is completely antisymmetric in the indices $[\beta\gamma\delta]$. The first condition will guarantee that there is no fourth propagating polarization (i.e., that $A_0$ remains non-dynamical), while the second condition is required for the three propagating polarizations to satisfy second order equations of motion. These conditions generalize to the multi-Proca interactions the existing results for the cases of scalar Horndeski and single generalized Proca. The fact that the structure of the interactions for the multi-Proca fields is more complex (involving both Lorentz and internal group indices) results in a more cumbersome relation for the required non-minimal couplings.
\section{Relation to generalized Proca}
\label{Sec:RelProca}
In this section we will compare the obtained interactions for the set of vector fields with an internal rotational symmetry with those of one single vector field. Let us remind here the corresponding interactions for the case of generalized Proca\footnote{Here $[\cdots]$ denotes the trace of the object inside the brackets.}
\begin{eqnarray}\label{vecGalcurv}
\mathcal L_2^{GP} & = & G_2(A_\mu,F_{\mu\nu}) \\
\mathcal L_3^{GP} & = & G_3[S] \nonumber\\
\mathcal L_4^{GP} & = & G_{4}R+G'_{4} \frac{ [S]^2-[S^2]}{4}
\nonumber\\
\mathcal L_5^{GP} & = & G_5G^{\mu\nu} S_{\mu\nu}-\frac{G'_{5}}{3}\frac{ [S]^3-3[S][S^2] +2[S^3]}{8} \nonumber\\
&+&\tilde{G}_5 \tilde{F}^{\mu\alpha}\tilde{F}^\nu_{\;\;\alpha} S_{\mu\nu} \nonumber \\
\mathcal L_6^{GP} & = & G_6L^{\mu\nu\alpha\beta} F_{\mu\nu}F_{\alpha\beta}
+\frac{G'_{6}}{2} \tilde{F}^{\alpha\beta}\tilde{F}^{\mu\nu}S_{\alpha\mu} S_{\beta\nu}\nonumber
\end{eqnarray} 
with $L^{\mu\nu\alpha\beta}\equiv\frac14\epsilon^{\mu\nu\rho\sigma}\epsilon^{\alpha\beta\gamma\delta} R_{\rho\sigma\gamma\delta}$ the double dual Riemann tensor, $G_{3,4,5,6}$ and $\tilde{G}_5$ arbitrary functions of $Y\equiv -\frac12 A^2$ and a prime denotes derivative with respect to $Y$. Now, we can easily extend the above interactions to the case of several vector fields. We will not impose any global symmetry a priori for the sake of generality. The first term $\LL_2^{GP}$ is trivially promoted to $\LL_2$ for multi-Proca and we will not discuss it further since it does not contain the novel derivative interactions we are interested in. The direct extension of the second term leads to
\be
\LL_3^{GP}\rightarrow G^a_3 S^{a\mu}_\mu
\ee
where $G^a_3$ is a set of Lorentz scalar functions of $A_\mu^a$. If $G^a_3$ is  an $SO(3)$ vector whose components are Lorentz scalar functions of $A_\mu^a$ (not necessary of $A_\mu^aA^{b\mu}\delta_{ab}$ since now more general terms are possible) then this term respects the global symmetry. The fact that we cannot construct such an $SO(3)$ vector with the fundamental objects at hand is another way of understanding our previous finding that we cannot construct the analogous of this term maintaining the global rotational symmetry, hence  
$\mathcal L_3^{GP} \to 0$.

In order to obtain non-trivial interactions at this order we need to use that the indices of $S$ in $\LL_3^{GP}$ can also be contracted by using the vector fields and the corresponding extension to the case of several vector fields will be
\be
G_{3,2} S_{\mu\nu} A^\mu A^\nu\rightarrow G^{abc}_{3,2} S^a_{\mu\nu} A^{b\mu} A^{c\nu}
\ee
with $G^{abc}_{3,2}$ an $SO(3)$ tensor with Lorentz scalar functions of the vector fields. Now we could choose $G^{abc}_{3,2}\propto \epsilon^{abc}$ to impose the rotational symmetry, but then we see that this would result in a vanishing interaction for symmetry reasons and we re-obtain our result that we need more than 2 vectors at this order to have non-trivial interactions which respect the global symmetry. 
Building $G^{abc}_{3,2}$ as an $SO(3)$ tensor in terms of $A_\mu^a$ (and possibly $\delta_{ab}$ and $\epsilon_{abc}$) we recover the rotationally invariant interactions obtained from our systematic construction. However, notice that $G^{abc}_{3,2}$ can be in general arbitrary (for instance, their components could simply be real parameters) so the global $SO(3)$ symmetry can be explicitly broken.

For $\LL_4^{GP}$, the direct extension to the case of a set of vector fields yields
\be
\LL_4^{GP}\rightarrow G_{4}R+G'_{4}\delta_{ab} \frac{ S^a_{\mu\nu}S^{b\mu\nu}-S^{a\mu}_\mu S^{b\nu}_\nu}{4}
\ee
where now $G_4$ is promoted to an arbitrary function of $Z\equiv-\frac12 A_\mu^a A^{a\mu}$. We could replace $\delta_{ab}$ for a more general object explicitly breaking the global symmetry. However, in that case the tuning with the non-minimal coupling to guarantee the absence of additional polarizations will be different.

For $\LL_5^{GP}$ we find the same difficulty as with $\LL_3^{GP}$ for a direct extension to the case of several vector fields. The extension of the terms corresponding to the vector Galileons $G_5$ and terms with pure intrinsic vector modes $\tilde{G}_5$ vanishes $\tilde{G}_5 \tilde{F}^{\mu\alpha}\tilde{F}^\nu{}_{\alpha} S_{\mu\nu}\to \tilde{G}_5 \tilde{F}^{a\mu\alpha}\tilde{F}^{b\nu}_{\;\;\alpha} S^c_{\mu\nu}\epsilon_{abc}=0
$. Again, this is due to the antisymmetry of $\epsilon_{abc}$ introduced to maintain the global symmetry. There is no direct extension of $\mathcal L_5^{GP}$ to the multi-Proca case that respects the global symmetry, hence $\mathcal L_5^{GP} \to 0$.

At next order in derivatives, we can again straightforwardly perform the direct extension of the single vector case yielding the non-trivial interactions for the multi-Proca case:
\be
\mathcal L_6^{GP} \to G_6L^{\mu\nu\alpha\beta} F^a_{\mu\nu}F_{a\alpha\beta}
+\frac{G'_{6}}{2} \tilde{F}^{a\alpha\beta}\tilde{F}^{\mu\nu}_aS^b_{\alpha\mu} S_{b\beta\nu}.
\ee
Notice that the contractions of the internal indices must be the one chosen in the above expression for the tuning with the non-minimal coupling to be the appropriate one. One could also promote the interaction with the alternative contraction of the internal indices $\tilde{F}^{a\alpha\beta}\tilde{F}^{b\mu\nu}S_{a\alpha\mu} S_{b\beta\nu}$, but this would require a different non-minimal coupling of the form $LFFAA$ and, thus, it would not correspond to a direct extension of the single vector field case, as we will show below.

We can then summarize the possible direct extensions of generalized Proca to the multi-Proca case with the global $SO(3)$ symmetry as
\begin{eqnarray}\label{vecMGalcurv}
\mathcal L_2^{MGP} & = & G_2(A_\mu^a,F^a_{\mu\nu}) \\
\mathcal L_4^{MGP} & = & G_{4}R+G'_{4}\delta_{ab} \frac{ S^a_{\mu\nu}S^{b\mu\nu}-S^{a\mu}_\mu S^{b\nu}_\nu}{4}
\nonumber\\
\mathcal L_6^{MGP} & = & G_6L^{\mu\nu\alpha\beta} F^a_{\mu\nu}F_{a\alpha\beta}
+\frac{G'_{6}}{2} \tilde{F}^{a\alpha\beta}\tilde{F}^{\mu\nu}_aS^b_{\alpha\mu} S_{b\beta\nu}. \nonumber
\end{eqnarray} 
Remarkably, the above family of Lagrangians is more restricted than the single vector field case, i.e., imposing the global symmetry substantially reduces the allowed interactions. This shows an expected resemblance with the multi-Galileon case where only two terms and one coupling constant are present \cite{multigalileon}. In our case, if we assume $G_2$ to be the simple Proca Lagrangian, the above family of interactions involve the coupling function $G_4$ (that would describe a multi-Galileon interaction in the decoupling limit) and $G_6$ that would describe the leading order interaction between the longitudinal and the transverse modes in the decoupling limit. Thus, going to the multi-Proca case only introduces one additional coupling with respect to the multi-Galileon case.

It is important to notice that our systematical construction presented in Section \ref{Sec:SystConst} also generated terms that cannot be achieved by extending any of the single Proca interactions. A class of such terms correspond to the interactions involving $\M_{a\dots}^{\mu_1\dots\nu_1\dots}$ in (\ref{Eq:FullLagrangian}) among which we have for instance
\begin{eqnarray}
\LL_4^{MGP}&=&\epsilon^{\alpha\beta\gamma\delta}\tilde{F}^a_{\alpha\lambda} S^{b\lambda}{}_\beta A^a_\gamma A^b_\delta.
\end{eqnarray} 
There is another class of interactions that cannot be directly obtained from single Proca which involve some contractions of the internal indices of the vector fields. Examples of those interactions are
\begin{eqnarray}
\mathcal L_3^{MGP} & = & S^{a\mu\nu} A_\mu^b A_\nu^d A_\alpha^c A^{e \alpha} \delta_{de} \epsilon_{abc} \nonumber\\
\mathcal L_5^{MGP} & = &\epsilon_{abc} A^a_\mu A^{\mu d} \tilde{F}_d^{\alpha\nu}\tilde{F}^{b\beta}_\nu S^c_{\alpha\beta}
\end{eqnarray} 
These terms will not require the introduction of non-minimal couplings because they are linear in $S^a_{\mu\nu}$. We also saw in Section \ref{Sec:SystConst}, that we can contract the internal indices in $[SSAA]$ in two different ways and both give terms that are not direct extensions of the single Proca case. However, we can use their resemblance to compute the required non-minimal couplings. Let us for instance consider the interactions in $\mathcal{L}_4^{(2A)}$ with the different contractions of the internal indices. In order to guarantee three propagating polarizations we would need to add different types of non-minimal couplings according to the contraction of the internal indices. In the following we show three different examples:
\begin{align}
&\mathcal L_4^{MGP}  \supset   A_\mu^aA_a^\mu A_\nu^b A_b^\nu R-A_\mu^a A_a^\mu\Big(S_{b\nu}^\nu S_{\alpha}^{b\alpha}-S_{b\alpha\nu}S^{b\alpha\nu}\Big)  \nonumber\\
 &+   A_\mu^aA_b^\mu A_\nu^b A_a^\nu R-A_\mu^a A^{b\mu}\Big(S_{a\nu}^\nu S_{b\alpha}^\alpha-S_{a\alpha\nu}S_b^{\alpha\nu}\Big)  \nonumber\\
 & +  A^{a\mu}A^{b\nu} A^\alpha_a A_b^\beta R_{\mu\nu\alpha\beta}-2A_\alpha^a A^{b}_\beta\Big(S_{a\nu}^\nu S^{b\alpha\beta}-\frac12S_{a\nu}^{(\alpha} S_b^{\beta)\nu}\Big) 
\end{align}
where each line gives an independent contribution. The same is true for the interactions in $\mathcal{L}_6$. If we contract the internal indices in a different way, the adjustment has to be done accordingly in the non-minimal coupling as, for instance, in
\begin{align}
&\mathcal L_6^{MGP} 
 \supset A_\rho^aA^{b\rho}  L^{\mu\nu\alpha\beta} F^a_{\mu\nu}F_{b\alpha\beta}
-4  \tilde{F}^{a\alpha\beta}\tilde{F}^{b\mu\nu} S_{a\alpha\mu} S_{b\beta\nu} \nonumber\\
&  +A_\rho^aA_a^\rho A_\sigma^b A_b^\sigma L^{\mu\nu\alpha\beta} F^c_{\mu\nu}F_{c\alpha\beta}
-8 A_\rho^aA_a^\rho \tilde{F}^{b\alpha\beta}\tilde{F}^{\mu\nu}_b S^c_{\alpha\mu} S_{c\beta\nu} \nonumber\\
 & +  A_\rho^aA^{b\rho} A_{a\sigma} A_b^\sigma L^{\mu\nu\alpha\beta} F^c_{\mu\nu}F_{c\alpha\beta}
-8 A_\rho^aA^{b\rho} \tilde{F}^{c\alpha\beta}\tilde{F}^{\mu\nu}_c S_{a\alpha\mu} S_{b\beta\nu}
 \end{align}
where again each line gives an independent contribution. Depending on what scalar functions have been considered in the general functions, the non-minimal couplings follow the corresponding structure of the internal indices. Their exact form can be extracted from the relations for $\Qm$ and $\Pm$ in section \ref{sec:GenForm}. As we mentioned in the introduction, in the case of Horndeski and generalized Proca interactions it was possible to construct terms with higher order equations of motion without altering the number of propagating degrees of freedom.  Some of these beyond Horndeski and beyond generalized Proca interactions arise if one simply promotes the flat space-time interactions to curved space-time without the relative tuning with the non-minimal couplings. Based on these known results, it is expected that beyond generalized multi-Proca interactions can be constructed by simply replacing the partial derivatives by covariant derivatives in the interactions obtained in section \ref{Sec:SystConst}. We leave the exploration of this direction for future works. 

\section{Cosmological applications}
\label{Sec:CosmApp}
In this section we will discuss the novel possibilities for cosmological applications offered by the interactions obtained in the previous sections. The distinction between terms with a global rotational symmetry and those without it is crucial for the possible cosmological solutions. Ultimately, the reason to impose such a symmetry is that it will allow configurations for the vector fields that break both spacetime and internal rotations (in addition to time translations and boosts) leaving a certain combination of the two unbroken, therefore offering the possibility of having isotropic solutions. However, even for interactions realizing an internal $SO(3)$ symmetry, the specific form of the interactions will lead to different cosmological scenarios, as we discuss in more detail below.

If the Lagrangian does not have an internal $SO(3)$ invariance, then the interactions are essentially equivalent to having a set of generalized Proca fields. These theories will allow isotropic solutions for configurations of the form $A^a_\mu=\phi^a(t)\delta^0_\mu$. If the interactions respect the internal $SO(3)$ symmetry, this configurations will break it along with boosts, but, in any case, spacetime rotations are preserved so the symmetry breaking pattern will be $SO(3,1)\to SO(3)$ or $SO(3,1)\times SO(3)_{\rm internal}\to SO(3)_{\rm spatial}$ if there is the internal invariance (which does not play any fundamental role for these configurations). This is analogous to the case of multi-scalar cosmological scenarios. Thus, we expect to have a phenomenology similar to the one that has already been explored in the literature where it has been shown the existence of de Sitter solutions as critical points \cite{cosmogenProca}. These solutions rely on the structure of the interactions that allow to have isotropic and homogeneous solutions supported by the temporal component of the vector, which is an auxiliary field. Although the temporal component does not propagate, it has a non-trivial effect on the cosmological evolution giving rise to a modified Friedman equation. In the case of several fields, we expect a similar phenomenology with the temporal components of the fields playing the role of several auxiliary fields.

Perhaps, more interesting at this stage are the interactions exhibiting a global $SO(3)$ symmetry because then we can have an additional isotropic source in the universe based on a configuration for the vector field of the form $A^a_{i}=A(t)\delta^a_i$. This type of configuration has also been considered in the literature for models of inflation, as in gauge-flation \cite{Maleknejad:2011jw} or chromo-natural inflation \cite{Adshead:2012kp}, and also as candidates for dark energy \cite{ArmendarizPicon:2004pm}. In this case, the fields configuration breaks both the spacetime and the internal rotations, but leaves an invariance under a linear combination of the two rotational groups, i.e., the symmetry breaking pattern in this case will be  $SO(3,1)\times SO(3)_{\rm internal}\to SO(3)_{\rm diagonal}$.

The existing models in the literature using the configuration discussed in the previous paragrapah do not consider the temporal components of the fields, either due to a non-abelian gauge invariance or because they were imposed to vanish by the field equations. However,  with the general interactions discussed in the previous sections, we can actually construct scenarios with a combination of the two field configurations aforementioned, i.e., we can have isotropic solutions where both the temporal components and the spatial parts are present and contribute to the cosmological evolution. For those solutions, the vector fields will acquire the form
\be
A_\mu^a= \phi^a(t) \delta_\mu^0 + A(t) \delta_\mu^a.
\label{conf1}
\ee
In this configuration, we have up to 4 scalar degrees of freedom, but the structure of the interactions will make the 3 components $\phi^a$ be auxiliary fields, so that only one dynamical dof is actually present. Since all the components will be tightly related by the structure of the interactions, their interplay will lead to novel and interesting scenarios for cosmological applications. However, the above configuration now breaks the two rotations and it is not clear that a linear combination survives. In fact, in general it is not the case and we need to add further restrictions to the interactions. This is not difficult to understand, as this configuration allows terms like $ T_{ij}\supset A^2\phi^a\phi^b\delta_{ai}\delta_{bj}$ in the energy-momentum tensor that give a non-vanishing anisotropic stress that can support the shear and, therefore, leads to a violation of isotropy\footnote{Although this new class of anisotropic cosmological solutions can be interesting in certain scenarios like novel setups for anistropic inflation, here we will focus on showing the possibility of having isotropic solutions.}. Alternatively, we can see that the above configuration does not contain any $SO(3)$ invariance inherited from the original symmetries. 
If we denote by $\omega_\mu{}^\nu$ and $J^a{}_b$ the Lorentz and internal generators respectively, then the total variation of the configuration in(\ref{conf1}) is given by
\begin{align}
\delta A^a_0&=J^a{}_{b}\phi^b\\
\delta A^a_i&=A\big(\omega_i{}^a+J^a{}_{i}\big)
\end{align}
where $\omega_i{}^a=\omega_i{}^\mu \delta_\mu^a$ and $J^a{}_{i}=J^a{}_{b}\delta^b_i$. From these variations we can clearly see the possibilities for the configuration to be isotropic. If $A=0$, then the rotations generated by $\omega_i{}^j$ are not broken. If $\phi^a=0$, the combination of internal and spatial rotations with $\omega_i{}^a+J^a{}_{i}=0$ remains as a symmetry (the aforementioned diagonal $SO(3)$). However, if both $\phi^a$ and $A$ are not vanishing, the field configuration does not possess any $SO(3)$ symmetry. Thus, the existence of isotropic solutions for the general configuration will further restrict the possible interactions. A straightforward way to achieve the desired features is to impose that the vector fields (without derivatives) only appear through the combination $A^2\equiv A_\mu^a A^\mu_a$. Effectively, this introduces one independent $SO(3)$ invariance for each Lorentz component that we could write as $J^a_{(\mu)b}$ so that the general configuration now can preserve a combination of the spatial rotations and the $SO(3)$ invariance of the spatial components, while the $SO(3)$ symmetry of the temporal components will be broken. Notice however that this symmetry does not commute with Lorentz boosts so that it would loose its internal character, but they do commute with spatial rotations. We should also mention that we do not need to impose this symmetry for interactions involving derivatives of the vector field. In fact, the only non-vanishing components of the strength fields for the general configuration in (\ref{conf1}) read $F^a_{0i}=\dot{A}\delta^a_i$ which is invariant under the diagonal $SO(3)$ and, therefore, we do not need to impose additional restrictions in the interactions where the $F$'s are only contracted among themselves, i.e., there is no mixing of both Lorentz and group indices of $F^a_{\mu\nu}$ and $A_\mu^a$. From this discussion, we conclude that Lagrangians in which the vector fields without derivatives only appear through $A^2$ and the derivative interactions are such that $F$'s and $S$'s do not mix their internal group indices, will allow isotropic solutions for the general configuration given in (\ref{conf1}).

In the following we will illustrate the different scenarios discussed above with a simple example. We will consider the usual spatially flat Friedman-Lema\^itre-Robertson-Walker metric
\be
\d s^2=-\d t^2+a^2(t)\d\vec{x}^2,
\ee
together with the most general Ansatz for the vector fields compatible with the symmetries of this spacetime which, according to our previous discussion, is given by the following configuration:
\begin{eqnarray}
A_\mu^a
 =
\begin{pmatrix}
\phi^1(t)&\phi^2(t)&\phi^3(t) \\
a(t)A(t)&0 &0 \\
0&a(t)A(t)&0 \\
0&0&a(t)A(t)
 \end{pmatrix}\,,
 \label{conf2}
\end{eqnarray}
 where we have introduced a normalising scale factor with respect to (\ref{conf1}) for later convenience. Let us now consider the Lagrangian
\begin{align}\label{Lag_cosm}
\LL=&-\frac14\Big(\alpha_1\delta_{ab}+\frac{\alpha_2}{M^2} A_{a\rho} A_b{}^\rho \Big)F_{\mu\nu}^aF^{b\mu\nu}\nonumber\\&-\frac12 M^2 A^2+\frac14 \Big(\lambda_1\delta_{ab}\delta_{cd}+\lambda_2\delta_{ac}\delta_{bd}\Big)A^a_\mu A^{b\mu} A_\nu^c A^{d\nu}
\end{align}
where $M$ is some mass scale and $\alpha_{1,2}$ and $\lambda_{1,2}$ are dimensionless parameters. This Lagrangian is simple enough to straightforwardly show the aforementioned features. We can see that the terms $\alpha_1$ and $\lambda_1$ satisfy the required conditions to have isotropic solutions, while the terms $\alpha_2$ and $\lambda_2$ do not. Thus, it is expected that the full configuration of the vector fields will allow for isotropic solutions only if $\alpha_2=\lambda_2=0$ (or it reduces to one of the other simpler configurations with either $A=0$ or $\phi^a=0$). This can be explicitly checked by computing the anisotropic stress, which is given by
\be
T^{ij}=\frac{1}{a^2}\left(\frac{\alpha_2}{M^2} E^2+\lambda_2 A^2\right)\phi^a\phi^b\delta_a^i\delta_b^j\quad\quad {\rm for}\quad i\neq j
\label{anisotropicstress}
\ee
for the off-diagonal components, while the difference of the diagonal components is
\be
T^{ii}-T^{jj}=\left(\frac{\alpha_2}{M^2} E^2+\lambda_2 A^2\right)\Big[(\phi^i)^2-(\phi^j)^2\Big]
\ee
where we have defined the electric field $E\equiv \dot{A}+H A$, with $H=\dot{a}/a$ the Hubble expansion rate. We then clearly see that the anisotropic stresses are automatically zero if $\alpha_2=\lambda_2=0$ i.e., if we turn off the terms that do not respect the additional invariance and, therefore, in that case we can have the full configuration of the vector fields while being compatible with isotropy. The other possibility is of course to switch off either $\phi^a$ or $A$ so that we have one of the more traditional configurations. If we look at the off-diagonal equations for the spatial components corresponding to $A^a{}_i$, we find
\be 
\frac{\alpha_2}{a^2M^2}\frac{\d}{\d t}\Big(a^2E\phi^a\phi_i\Big)-\lambda_2A\phi^a\phi_i=0.
\label{Eqoffdiagonal}
\ee
Again, this equation trivializes for $\alpha_2=\lambda_2=0$ and, thus, we are left only with the equations for $\phi^a$ and the diagonal part of $A^a_i$. In addition to the anisotropic stresses there could be another source of isotropy violation provided by $T^{0i}$, which is given by
\be
T^{0i}=-\frac{A}{a}\left[M^2+(\lambda_1+\lambda_2)\Phi^2-(3\lambda_1+\lambda_2)A^2+\frac{\alpha_ 2E^2}{M^2} \right]\phi^i.
\ee
This expression persists even in the case of $\alpha_2=\lambda_2=0$. However, if we look at the equation for $\phi^a$, we find
\be
\left[M^2+(\lambda_1+\lambda_2)\Phi^2-(3\lambda_1+\lambda_2)A^2+\frac{\alpha_ 2}{M^2} E^2\right]\phi^a=0
\ee
and, therefore, $T^{0i}$ vanishes on-shell. Notice that this result does not rely on the absence of the terms $\alpha_2$ and $\lambda_2$ violating the additional invariance and it is due to the relation $T^{0i}=-A^{ai}\delta\LL/\delta\phi^a$ that holds for a large class of theories. In practice, this means that, quite generally, we only need to guarantee the absence of anisotropic stresses from the interactions. This quite general result was also noticed in \cite{Jimenez:2009ai} for more traditional vector-tensor theories.

For the theory with $\alpha_2=\lambda_2=0$, the non-trivial field equations read
\begin{align}
&\phi^a\Big[ \lambda_1(\Phi^2-3A^2)+M^2\Big]=0, \\
&\dot{E}+2HE+\Big[M^2+\lambda_1(\Phi^2-3A^2)\Big]A=0,
\end{align}
where we have additionally set $\alpha_1=1$ to match the normalization of a Maxwell field. Thus, we see that the effect of $\phi^a$ is to cancel the impact of the potential on $A$ so that it evolves as a standard Maxwell field with $E\propto a^{-2}$. This is general if we replace the quartic potential by an arbitrary potential of the form $V(A^2)$. The value of $\Phi^2$ will be such that the field remains at the minimum of the potential (more precisely, it will make $V'=0$) while $A$ will evolve as a usual gauge field. This means that we will effectively have three identical Maxwell fields giving a radiation-like contribution plus a cosmological constant given by the value of the potential at the minimum. For our simple case, the energy density after inserting the solution for $\Phi^2$ reads
\be
\rho=\frac{M^4}{4\lambda_1}+\frac{3}{2}E^2
\ee
where we clearly see the effective cosmological constant plus the radiation component. This is not a general result, but a consequence of the simple Lagrangian considered, even if an arbitrary potential is introduced. For more general interactions including non-minimal couplings and genuine novel derivative self-interactions involving $S_{\mu\nu}^a$, integrating out the fields $\phi^a$ will impact the evolution of $A$ and the energy-density in a much richer way. 

So far, we have been interested in interactions preventing the appearance of anisotropic stresses for our general vector fields configuration. However, we can alternatively cancel ($\ref{anisotropicstress}$) by looking for solutions to the field equations with $\alpha_ 2 E^2+\lambda_2 A^2=0$, but this is not, in general, compatible with the vector fields equations. In this potential branch of solutions, we end up with an overdetermined system of equations and, thus, the existence of non-trivial solutions is not guaranteed. It might happen that for some special interactions, the eventual overdetermined system of equations admit non-trivial solutions. In any case, these will represent very particular cases and we find more natural to consider interactions with the additional invariance permitting isotropic solutions with the general configuration given in (\ref{conf2}). 

From the interactions considered throughout this paper, we can easily identify a subset of terms allowing for the general configuration of the vector fields given in (\ref{conf1}) while respecting isotropy. Such terms can be expressed as $\LL=\sum_i \sqrt{-g} \LL_i$ with
\begin{align}
\LL_2=&G_2(A^2,F^a_{\mu\nu})\;,\nonumber\\
\LL_4=&G_{4}R+G'_{4}\delta_{ab} \frac{ S^a_{\mu\nu}S^{b\mu\nu}-S^{a\mu}_\mu S^{b\nu}_\nu}{4}\;,\nonumber\\
\LL_6=& G_6L^{\mu\nu\alpha\beta} F^a_{\mu\nu}F^a_{\alpha\beta}
+\frac{G'_{6}}{2} \tilde{F}^{a\alpha\beta}\tilde{F}^{a\mu\nu}S^b_{\alpha\mu} S^b_{\beta\nu}.
\end{align}
This is the direct generalization of the single field case to the case of multi-Proca with the global $SO(3)$ symmetry discussed in the previous section with the only difference that $A_\mu^a$ can only enter all the $G_i$ functions through $A^2$, while the dependence on $F_{\mu\nu}^a$ of $G_2$ remains arbitrary. Notice that because of the restriction of its dependence on $A_\mu^a$, there cannot be mixing of both Lorentz and internal group indices between $A_\mu^a$ and $F_{\mu\nu}^a$. The resulting theory is remarkably simple with only three terms so that it gives an appealing framework for novel cosmological scenarios.

To conclude this section, we will comment on another potential branch of isotropic solutions. It was shown in \cite{Cembranos:2012kk} that general vector fields (both abelian and non-abelian) with a potential can also provide isotropic solutions supported by oscillating fields with vanishing temporal components and arbitrary configuration for the spatial components. This is achieved by averaging over several oscillations when the fields oscillate with a frequency much higher than the Hubble expansion rate. This result was also extended to fields of arbitrary spin \cite{Cembranos:2013cba}. It would be interesting to explore this class of solutions for our more general interactions which would provide a cosmological scenario different from the one we consider here.

\section{Discussion}
\label{Sec:Conc}
In this work we have constructed derivative self-interactions for a set of vector fields following a systematic approach based on the totally antisymmetric Levi-Civita tensor. We have considered interactions with the schematic form $[\epsilon\epsilon(\partial A)^m A^{2n}]$ and then we have proceeded in increasing number of derivatives $m$ and fields without derivatives $n$. In order to limit the number of possible interactions and motivated by the potential cosmological applications, we have focused on interactions with a global $SO(3)$ symmetry. After constructing the interactions we have discussed the relation with the derivative self-interactions for one single vector field of the generalized Proca theories. The obtained terms can be broadly classified into those which represent a direct extension of the generalized Proca interactions and those which are genuinely novel and do not exist for the case of one single vector field. 

In addition to building the interaction terms we have discussed the potential cosmological applications. In order to do that, we have considered the different configurations for the vector fields that could be compatible with an isotropic universe. There are two simple configurations that are compatible with isotropy and which correspond to those already considered in the literature. This first class corresponds to the vector fields having only temporal components. For this configuration the global $SO(3)$ invariance is not required to have isotropic solutions since spatial rotations are not broken. The second class of solutions corresponds to a configuration where the temporal components vanish while the spatial components are given by $A^a_i=A\delta^a_i$. In this case the global $SO(3)$ invariance is crucial because even if spatial rotations are broken, a combination of internal and spatial rotations remains unbroken and allows for isotropic solutions. Finally, we have discussed the existence of more general isotropic solutions where the fields configuration is a combination of the previous ones, namely $A^a_\mu=\phi^a\delta^0_\mu+A\delta^a_\mu$. However, having isotropic solutions for this configuration is not as obvious and, in fact, we need additional restrictions in the interactions. We have argued how to restrict the interactions in order to allow for isotropic solutions with the general configuration and we have shown these features for a simple Lagrangian. Thus, these interactions open new possibilities for cosmological scenarios not considered so far in the literature. Here we have contented ourselves with the discussion and illustration of the general structure of the solutions and we will leave a more thorough analysis of cosmological solutions for a subsequent work.

\acknowledgments 
We thank Federico Piazza for very useful discussions. JBJ  acknowledges  the  financial  support  of
A*MIDEX project (n ANR-11-IDEX-0001-02) funded by
the  "Investissements d'Avenir" French Government pro-
gram, managed by the French National Research Agency
(ANR),  MINECO  (Spain) projects
FIS2014-52837-P and Consolider-Ingenio MULTIDARK
CSD2009-00064. L.H. acknowledges financial support from Dr. Max R\"ossler, the Walter Haefner Foundation and the ETH Zurich Foundation.

\end{document}